\documentclass[12pt]{article}
\usepackage[T1]{fontenc}
\usepackage[latin1]{inputenc}
\usepackage{amsmath}
\usepackage{amssymb}
\usepackage{graphicx}

\begin{document} 

\title{Scattering of Classical and Quantum Particles by Impulsive Fields}

\author{
    Herbert Balasin\thanks{hbalasin@tph.tuwien.ac.at}\\
    Institut f\"ur \\
    Theoretische Physik\\
    TU-Wien
  \and
    Peter C. Aichelburg\thanks{aichelp8@univie.ac.at}\\
    Fakult\"at f\"ur Physik\\
    Gravitationsphysik\\
    Universit\"at Wien}

\maketitle
\vspace{2cm}
\begin{abstract}

We investigate the scattering of classical and quantum particles in impulsive backgrounds fields. These fields model short outbursts of radiation propagating with the speed of light.
The singular nature of the problem will be accounted for by the use of Colombeau's generalized function which however give rise to ambiguities. It is the aim of the paper to show that these ambiguities can be overcome by implementing additional physical conditions, which in the non-singular case would be satisfied automatically. As example we discuss the scattering of classical, Klein-Gordon and Dirac particles in impulsive electromagnetic fields.

\end{abstract}
\newpage
\section*{Introduction}

In the following we consider the scattering of  classical (point-like) as well as quantum
(waves) particles by impulsive  background fields. That is to say, the particles interact with
a field that is solely concentrated on a null hyperplane described by a delta-like singularity.
The physical motivation is to model the behavior of particles affected by
extreme short outbursts of radiation such as observed in supernovae
explosions, gamma-ray bursts or the fields of ultra-short laser pulses
produced in the laboratory.
Since, from the spacetime point of view,  the particles move freely ``above''
and ``below'' the pulse-hyperplane the solution of the equations of motion
is turned into a matching problem for free solutions.

The mathematical price for this simplified physical description comes in
the form of non-linear operations performed on distributional
objects. The adequate framework is provided by the algebra of new
generalized functions $\mathcal{G}$ of Colombeau \cite{Col}. It circumvents the
Schwarz impossibility result, that claims the non-existence of a (differential)
algebra extending the continuous functions and containing the distributions, by
requiring only the $C^\infty$-functions to be a sub-algebra.

Early work by DeVega and Sanchez \cite{dVSa} and Lousto and Sanchez \cite{LoSa}
discuss the scattering of Klein-Gordon and Dirac fields in a special class of impulsive
gravitational backgrounds. These geometries can be obtained from the ultra-relativistic limit
of black hole space-times  (AS-geometries \cite{AS} and generalizations
thereof \footnote{For an initial value approch to impulsive gravitational waves cf. \cite{RoLu}}).
The authors notice a particular regularization dependence of their result.

On the other hand Kunzinger and Steinbauer \cite{KuSt} have investigated the
behavior of geodesics in general impulsive pp-wave backgrounds via
rigorously embedding the equations into the Colombeau algebra. They show that
a (unique) solutions to the geodesic as well as the geodesic deviation
equation exist in $\mathcal{G}$ which possess a reasonable macroscopic, that
is distributional, ``shadow''. These results are in accordance with earlier
work by Balasin \cite{Ba1}.

In the present paper we follow a similar strategy as in  \cite{Ba1}
whence extended to the field context. Distributional equality will be replaced
by association which is the corresponding notion in $\mathcal{G}$. We find
that in spite of the singular character the nonlinear operations yield
associated distributional objects containing, however, finite undetermined
quantities, which is to be expected \cite{Col}. In order to define their
values from a physical point of view we make use of conservation
laws. These would follow in the smooth context via nonlinear operations which,
in general however, break association. We believe that this method provides a
systematic way to decide upon the ``regularization dependence'' in \cite{dVSa}
without relying on the heavy machinery used in \cite{KuSt} (in fact being
closer to the approach used in \cite{Col})
In sec. 1 we briefly recall the definition of the Colombeau algebra and
discuss the properties of association.
As a warm up, we start in sec. 2  by considering  a  particle subject to  a potential acting only at a single instant of time.
Already in the Newtonian (classical) case, care must be taken in handling the non-linearities coming from the potential, but the solution is uniquely determined. In contrast,  the corresponding Schr\"odinger problem, in sec. 3,  leads to an undetermined quantity. The reason is the association of the product of the $\delta$ term in the potential with the $\theta$ functions of the matched free solutions. Rather than stipulating its value by hand, we require the conservation of probability through the pulse, which does not follow from the equation since it involves non-linear operations  which in general break distributional equality, i.e. association. From this we find via a Cayley-transform the uniquely defined transition amplitude.
In a next step, in sec. 4, we consider a classical particle in an electromagnetic pulse
and the corresponding Klein-Gordon equation as its quantum version (sec. 5) . Although the
situation is more involved the strategy is precisely the same as in the
Newton-Schr\"odinger case. Here already the classical problem gives rise to an ambiguity which can be  fixed
by requiring that the length of the tangent vector to be preserved.

The case of the Klein-Gordon field is still more involved because it leads to two undetermind constants, which at first sight seems hopeless for obtaining a unique solution.
However, we show that by the  physical requirement that the Klein-Gordon
current is conserved across the pulse, the two constants are determined and thus the matching is unique.
Finally we focus on relativistic particles with spin i.e.  we consider the impact of an electromagnetic pulse on Dirac-particles.  Again undefined quantities (constants) arise. It is natural to impose the conservation of the Dirac current across the pulse which then gives the unique transition amplitude.

\section{The method}

The Colombeau algebra $\mathcal{G}$ consists of one-parameter families
of $C^\infty$ functions, $(f_\epsilon(x))_{(0<\epsilon<1)}$ subject to certain
growth-conditions in $\epsilon$. Its elements may be
thought of as being regularizations of distributional (and even more
singular) objects. Distributions form a linear subspace of $\mathcal{G}$. This
subspace is not canonical in the same sense as $SU(2)$ is not a {\em canonical}
subgroup of $SL(2,\mathbb{C})$. In particular this means that there are many
different ``$\delta$-functions'' in $\mathcal{G}$. This property is reflected
by an equivalence relation on the algebra called association and denoted by $\approx$
\begin{equation}
  \label{eq:assoc}
(f_{\epsilon}(x))\approx(g_{\epsilon}(x))\qquad iff\qquad \lim_{\epsilon\to0}\int d^nx(f_{\epsilon}-g_{\epsilon})(x)\varphi(x)=0\qquad\forall\varphi\in C_{0}^{\infty}(\mathbb{R}^n).
\end{equation}
Objects in the same equivalence class may differ in their micro-aspect.
That is to say although they are in general different objects in
$\mathcal{G}$, they all correspond (if it exists) to the same
distribution. In this regard we may think of association as a kind of
coarse-graining of $\mathcal{G}$. Distributionally well-defined, i.e. linear,
operations have well-defined analogs in $\mathcal{G}$, which are compatible with their macro-aspect, meaning they do
not break association
\begin{equation}
  (f_\epsilon) \approx (g_\epsilon)\quad\left\{
  \begin{array}{r@{\quad\Rightarrow\quad}l}
    &f\cdot(f_\epsilon) \approx f\cdot(g_\epsilon)\\
    &(\partial_\alpha f_\epsilon) \approx (\partial_\alpha g_\epsilon)
  \end{array}\right.
\end{equation}
On the other hand, non-linear operations on different representatives of an
association-class do in general break association. This means that upon
non-linear operations different micro-aspects may get magnified to the macro-level.
A simple, nevertheless important, example is given by
\begin{equation}
\theta\cdot \delta \approx A \delta
\end{equation}
where $A$ denotes a constant. This simply states that $A$ is the result of the relative micro-aspects of the two
elements of $\mathcal{G}$ that are associated to $\theta$ and $\delta$ respectively.

As a special case we have
\begin{equation}
  \label{eq:ththpr}
 (\theta^2 )'\approx\theta\cdot \theta'\approx \frac{1}{2}\delta
\end{equation}
or more generally by
\begin{equation}
  \label{eq:thnthpr}
  \theta^n \theta'\approx \frac{1}{n+1}\delta.
\end{equation}
These results follow from $\theta^n\approx\theta$
and the compatibility of association with differentiation.
In both the above cases the constant is determined regardless of the
representative $\theta$. This reflects the fact that the relative
micro-aspect between $\theta$ and $\theta'$ and $\theta^n$ and $\theta'$
respectively is independent of the representative. Notice that all $\theta^n$ are again
$\theta$-functions, i.e. are associated to the $\theta$-distribution. However,
their different micro-aspects relative to $\theta'$ a representative of
$\delta$ gets magnified to the macro-level.

\section{Newtonian particle}
Having prepared the stage, let us apply the formalism to the simple-most  classical system: a particle under the influence of a potential acting only at an instant of time, described by the Newtoniam equation of motion.
\begin{equation}
  m\ddot{x}^i(t)+\delta(t)\partial_i V(x^m(t))\approx0.\label{cleom}
\end{equation}
We have chosen weak equality because the $\delta$ term representing
the idealized action of the force during the shortest possible period
of time. Since the force acts only at $t=0$
the trajectory is given by
\begin{eqnarray*}
x^i(t)=\theta_+(t)x^i_+(t)+\theta_-(t)x^i_-(t), &  & x^i_+(t),x^i_-(t)\in C^{\infty}(\mathbb{R})\\
\theta_+(t)+\theta_-(t)=1,
\end{eqnarray*}
We require $x^i_-(t)$ and $x^i_+(t)$ to be solutions of the free equations
of motions before and after the pulse respectively.

Upon insertion into (\ref{cleom}) gives
\begin{multline*}
m(x^i_+(0)-x^i_-(0))\delta'(t)+m(\dot{x}^i_+(0)-\dot{x}^i_-(0))\delta(t)+\\
\theta_+(t)m\ddot{x}^i_+(t)+\theta_{-}(t)m\ddot{x}^i_-(t)+\delta(t)\partial_i
V(x^m(t))\approx0.
\end{multline*}

The last term contains, via
the association process, undetermined constants as pre-factors of
$\delta(t)$, i.e. $\delta(t)\partial_iV(x^m(t))\approx C_i\delta(t)$.

In a first step,  multiplication with the $C^{\infty}$ function $t$ ensures the vanishing of all the terms in (\ref{cleom}) except the $\delta'$ term, since $t\delta'$ is not $\approx 0$
tells us that the coefficient has to vanish, i.e. $x^i_+(0)=x^i_-(0)$.
This then in turn determines the prefactor to be $C_i=\partial_iV(x^m(0))$.

So we are only left with the $\delta$ function coefficient
\begin{equation*}
m\dot{x}^i_+(0)-m\dot{x}^i_-(0)=-\partial_i V(x^m(0)),
\end{equation*}
where $x^i(0)$ has now a well-defined meaning. The junction condition
has a simple physical interpretation as mapping the $t=0$ conditions of
for $x_-$ onto the $t=0$ condition for $x_+$, i.e.
\begin{equation}
  \label{eq:in_map}
(x^i_-(0),\dot{x}^i_-(0))\mapsto(x^i_+(0),\dot{x}^i_+(0))=(x^i_-(0),\dot{x}^i_-(0)-\frac{1}{m}\partial_i V(x^m(0))).
\end{equation}
As expected, the freely moving particle gets a kick at $t$ = 0, thereby changing its velocity according to the applied force.

\section{Schr\"odinger particle}

Quantization of the above system is described by
the corresponding  time-dependent, Schr\"odinger equation

\begin{equation}
\label{schroe}
i\dot{\psi}(x,t)+(-\frac{\partial^2}{2m}+\delta(t)V(x))\psi(x,t)\approx0
\end{equation}

 Proceeding in the same way as with the classical system we combine
two solutions before and after $t=0$, i.e.
\begin{equation}
\label{ansatz}
\psi(x,t)=\theta_+(t)\psi_+(x,t)+\theta_-(t)\psi_-(x,t).
\end{equation}
where we assume that $\psi_-(x,t)$ and $\psi_+(x,t)$ satisfy the free Schr\"odinger equation.

Inserting (\ref{ansatz}) into (\ref{schroe}) we are left with
\begin{equation*}
i\delta(t)(\psi_+(x,0)-\psi_-(x,0))\approx\delta(t)V(x)(A\psi_+(x,0)+
    (1-A)\psi_-(x,0)).
\end{equation*}

The above relation made implicit use of $\theta\cdot\delta\approx A\delta$,
which expresses our ignorance about the microscopic relation between
$\theta$ and $\delta$. As in the classical regime this relates the data $\psi_-(x,0)$
before the shock to the data $\psi_+(x,0)$ after the shock
\begin{equation}
\psi_+(x,0)=\frac{1-i(1-A)V(x)}{1+iAV(x)}\psi_-(x,0).\label{datamap}
\end{equation}
There is however an important difference, that manifests itself in
the appearance of the arbitrary constant $A$. It signals that we have
oversimplified the physical description of the system by making it too
singular. Following \cite{Col} the mathematical  description would need further specification. It is precisely additional physical input that allows to determine $A.$ For general smooth solutions
of the Schr\"odinger equation we have conservation of the probability
current. However due to the weak nature of our equation current conservation
is no longer one of its consequences. In order to preserve the physical interpretation
of the Schr\"odinger equation we have
\begin{eqnarray*}
  \dot{\rho}+\partial_i j^i\approx 0, &  & \rho=\bar{\psi}\psi,\qquad
  j^i=\frac{1}{2mi}(\bar{\psi}\partial_i\psi-
  \psi\partial_i\bar{\psi}),\\
  \rho & \approx & \theta_+\bar{\psi}_+\psi_++\theta_-\bar{\psi}_-\psi_-,\\
  j^i & \approx &\frac{1}{2mi}(\theta_+(\bar{\psi}_+\partial_i\psi_+-
  \psi_+\partial_i\bar{\psi}_+)+\theta_-(\bar{\psi}_-\partial_i\psi_--
  \psi_-\partial_i\bar{\psi}_-)),\\
  \dot{\rho}+\partial_i j^i & \approx &  \delta(\bar{\psi}_+\psi_+-
  \bar{\psi}_-\psi_-).
\end{eqnarray*}
Using (\ref{datamap}) has the immediate consequence
\begin{equation*}
\bar{\psi}_+\psi_+=|\frac{1-i(1-A)V}{1+iAV}|^2\bar{\psi}_-\psi_-.
\end{equation*}
Therefore probability-current conservation is only achieved if the
pre-factor in the last equation has unit length, thereby fixing $A$
to be $1/2$. With this, the data below the pulse are mapped uniquely via a Cayley-transform of the reduced potential into the data
above the pulse.
\begin{equation*}
  \psi_+(x,0)=\frac{1-\frac{i}{2}V(x)}{1+\frac{i}{2}V(x)}\psi_-(x,0).
\end{equation*}

We mention that our formalism also
works in the more common situation of a spatially "impulsive" potential
$V(x^m)=\delta(nx)\tilde{V}(\tilde{x}^m)$ where similar ambiguities arise, but
do not contribute to the result as for the  Newtonian particle.

\section{Lorentz particle}
Turning to a relativistic setting it is natural to consider disturbances that 
travel with the fundamental velocity i.e. along null, rather than on  $t=const$ surfaces.
Therefore, we look at the scattering by impulsive electromagnetic fields which
are completely concentrated on a null hyperplane.
The vector potential $A_{a}$ and the field-strength $F_{ab}$ take
the form
\begin{equation}
A_{a}=f(px,\tilde{x}^m)p_{a},\,\, F_{ab}=2\tilde{\partial}_{[a}fp_{b]},\qquad f(px,\tilde{x}^m)=\delta(px)\tilde{f}(\tilde{x}^m)\label{empot}
\end{equation}
 where $p^{a}$ denotes (a covariantly constant) null vector-field and
$\tilde{x}^i$ denotes the spacelike coordinates of the two-dimensional
subspace orthogonal to $p^{a}$ and a conjugate null direction $\bar{p}^{a}$.
Here we follow the coordinate free notation of Penrose \cite{Pen}. However, if one introduces the coordinates
 $(u,v, \tilde{x}^m)$ in Minkowski space

\begin{equation}
ds^{2}= 2dudv -d\tilde{x}^md\tilde{x}^m
\end{equation}
and chosing  $p = {\partial}_{v}$ and $\bar{p}= {\partial}_u$, then $px = u$ and $\bar{p}x = v$.

$F_{ab}$ satisfies the vacuum equations provided
\begin{equation}
 \Delta_2 \tilde{f}(\tilde{x}^m) = 0
\end{equation}

The motion of test-particles $x^{a}(s)$ is described by the Lorentz-force law
\begin{eqnarray*}
m\ddot{x}^{a}+eF^{a}\,_{b}\dot{x}^{b} & = & 0,\\
m\ddot{x}^{a}+e((p\dot{x})\tilde{\partial}^{a}f-(\dot{\tilde{x}}\tilde{\partial})fp^{a}) & = & 0,
\end{eqnarray*}
which becomes upon decomposition with respect to $p^{a},\bar{p}^{a}$
and their orthogonal complement
\begin{eqnarray}
p\ddot{x} & = & 0,\nonumber \\
\bar{p}\ddot{x}+\frac{e}{m}(\dot{\tilde{x}}\tilde{\partial})f & = & 0,\nonumber \\
\ddot{\tilde{x}}+\frac{e}{m}(p\dot{x})\tilde{\partial}f & = & 0,\label{emshock1}
\end{eqnarray}
where we have suppressed the indices in the two-dimensional (tilde)
part and made use of the normalization $p\cdot\bar{p}=1$. The first
equation of (\ref{emshock1}) tells us that $px$ may be chosen as
an ``affine'' parameter for the trajectory ($(px)\dot{}=\alpha$) unless we consider motion
within a hyperplane orthogonal to $p^{a}$. Taking into account the
impulsive nature of the profile $f$ the equation will be considered
as weak equality within the Colombeau algebra
\footnote{Here and in the following notation like
  $(\bar{p}x)(px)\equiv(\bar{p}x)(u)$ denotes the dependence of $\bar{p}x$ and
  similar expressions on the affine parameter $px=u$} 
\begin{eqnarray}
(\bar{p}x)''(px)+\frac{e}{\alpha m}\delta(px)
(\tilde{x}^i{}'(px)\tilde{\partial}_i\tilde{f}(\tilde{x}^m(px)) & \approx &
0,\nonumber\\
\tilde{x}^i{}''(px)+\frac{e}{\alpha m}\delta(px)\tilde{\partial}_i\tilde{f}(\tilde{x}^m(px)) & \approx & 0.\label{emshock2}
\end{eqnarray}

Since the electromagnetic field is completely concentrated on the
plane $px=0$ the particle moves freely ``above'' and ``below''
the pulse, i.e.
\begin{eqnarray*}
(\bar{p}x)''(px) & = & \theta_{+}(px)(\bar{p}x_{+})(px)+\theta_{-}(px)(\bar{p}x_{-})(px),\\
\tilde{x}^i{}''(px) & = & \theta_{+}(px)\tilde{x}_{+}^i(px)+
\theta_{-}(px)\tilde{x}_{-}^i(px).
\end{eqnarray*}
The second equation of (\ref{emshock2}) is identical to that for
the Newtonian particle. Therefore the junction conditions become
\begin{eqnarray*}
 &&\tilde{x^i}_{+}(0)=\tilde{x^i}_{-}(0),\\
 &&\tilde{x^i}_{+}'(0)=\tilde{x^i}_{-}'(0)-\frac{e}{\alpha m}\tilde{\partial}_i\tilde{f}(\tilde{x}^m(0)).
\end{eqnarray*}
Let us now take a closer look a the first equation of (\ref{emshock2}).
\begin{multline*}
((\bar{p}x_{+})(0)-(\bar{p}x_{-})(0))\delta'(px)+((\bar{p}x_{+})'(0)-
(\bar{p}x_{-})'(0))\delta(px)+\\
\frac{e}{\alpha
  m}\delta(px)(\theta_{+}'(px)(\tilde{x}_{+}^i(px)-\tilde{x}_{-}^i(px))+\\
\theta_{+}(px)\tilde{x}_{+}'^i(px)+\theta_{-}\tilde{x}_{-}'^i(px))\tilde{\partial}_i\tilde{f}(\tilde{x}^m(px))  \approx  0.
\end{multline*}
Due to the appearance of products like $\theta'(px)\cdot\delta(px)$ the
above expression makes only sense within the algebra. Multiplication
with $px$ and taking into account the junction conditions for $\tilde{x}^i(px)$
along the lines of the Newtonian particle, shows that the coefficient of the $\delta'(px)$ term has to vanish separately,
i.e.
\begin{equation*}
  (\bar{p}x_{+})(0)=(\bar{p}x_{-})(0).
\end{equation*}
From the remaining expression we obtain
\begin{equation*}
  (\bar{p}x_{+})'(0)-(\bar{p}x_{-})'(0)+\frac{e}{\alpha
    m}(A\tilde{x}_{+}'^i(0)+(1-A)\tilde{x}_{-}'^i(0))
  \tilde{\partial_i}\tilde{f}(\tilde{x}^m(0))=0,
\end{equation*}
where, as before, the (remaining) constant $A$ arises from $\theta(px)\cdot\delta(px)\approx A\delta(px)$.
Let us pause for a moment and compare our results with the Newtonian
case. Although we have obtained the arbitrary constant $A$ in very much
the same way, this arbitrariness already appears at the classical level.
In order to fix this constant we will invoke a consequence of the
equation of motions for smooth solutions, namely the fact that the
length of the tangent vector remains constant along the trajectory

\begin{multline*}
-2(\bar{p}x)'(px)+(\tilde{x}'(px))^{2}  \approx  const,\\
-2(\theta_{+}(px)(\bar{p}x_{+})'(px)+\theta_{-}(px)\bar{p}x_{-}'(px))+
\left(\theta_{+}'(px)(\tilde{x}_{+}(px)-\tilde{x}_{-}(px))+\right.\\
\left.\theta_{+}(px)\tilde{x}_{+}'(px)+\theta_{-}(px)\tilde{x}_{-}'(px)\right)^2  \approx  const,\\
\theta_{+}(px)(-2\bar{p}x_{+}'(px)+(\tilde{x}_{+}'(px))^2)+
\theta_{-}(px)
(-2\bar{p}x_{-}'(px)+(\tilde{x}_{-}'(px))^{2})  \approx  const,
\end{multline*}
Since differentiation does not break association, we have
\begin{multline*}
(-2\bar{p}x_{+}'(0)+(\tilde{x}_{+}'(0))^{2}+2\bar{p}x_{-}'(0)
-(\tilde{x}_{-}'(0))^{2})\delta(px)  \approx  0,\\
2\frac{e}{\alpha m}(\tilde{x}_{-}'^i(0)-A\frac{e}{\alpha
  m}\tilde{\partial}_i\tilde{f})\tilde{\partial}_i\tilde{f}+(\tilde{x}_{-}'(0)-
\frac{e}{\alpha m}\tilde{\partial}\tilde{f})^{2}-(\tilde{x}_{-}'(0))^{2}
 =  0.
\end{multline*}
This condition is only satisfied if $A$ is taken to be $1/2$.
Thus, summing up and denoting the "jump"  at $px = 0$ by [  ], we have:
\begin{eqnarray*}
  \left[ \tilde{x}^i \right]&=& 0\\
  \left[ \tilde{x}^i{}'  \right]&=& -\frac{e}{\alpha m}\tilde{\partial}_i\tilde{f}(\tilde{x}^m(0))\\
  \left[ \bar{p}x \right]&=& 0\\
  \left[ \bar{p}x' \right]&=& -\frac{e}{\alpha m}(\tilde{x}'_-(0)\tilde{\partial}\tilde{f}(\tilde{x}^m(0))) +
                (\frac{e}{\alpha m})^2(\tilde{\partial}\tilde{f}(\tilde{x}^m(0)))^2
\end{eqnarray*}

\section{Klein-Gordon particle}

Quantization gives rise to
\begin{eqnarray}
  \label{eq:KG}
&&(\eta^{ab}\hat{P}_a\hat{P}_b-m^2)\Phi=0\nonumber\\
&&\text{where $\hat{P}_a$ is given by}\quad \hat {P}_a=(\hat{p}_a-eA_a)\quad\hat{p}_a=\frac{1}{i}\partial_a\nonumber\\
&&(\eta^{ab}(\partial_a -ieA_a )(\partial_b -ieA_b)+m^2)\Phi=0.
\end{eqnarray}
Using the specific form of the potential (\ref{empot}) and taking
into account the lightlike nature of $p^a$, the above expression
simplifies to
\begin{equation}
  \label{eq:ppkg}
(\partial^2 + m^2 -2ief(p\partial))\Phi=0.
\end{equation}
The standard decomposition $\Phi=\theta_+\Phi_++\theta_-\Phi_-$,
resulting from the impulsive nature of $f$, i.e. $f=\delta(px)\tilde{f}$
yields upon insertion into (\ref{eq:ppkg})

\begin{eqnarray*}
2((p\partial)\Phi_+ - (p\partial)\Phi_- )\delta+\theta_+ (\partial^2 + m^2)\Phi_+
  + \theta_-(\partial^2 + m^2)\Phi_- +\\
-2ie\delta\tilde{f}(A(p\partial)\Phi_+ + (1-A)(p\partial)\Phi_- ) & \approx & 0
\end{eqnarray*}
where in order to have well-defined products of singular quantities
(\ref{eq:ppkg}) has been promoted to a weak statement within the Colombeau
algebra. Since $\Phi_+ ,\Phi_- $ satisfy the free Klein-Gordon
equation ``above'' and ``below'' the pulse respectively, we find
for the mapping from the final data of $\Phi_- $ to the initial
data of $\Phi_+ $
\begin{equation}
  \label{eq:jckg}
  (p\partial)\Phi_+ - (p\partial)\Phi_- =
  ie\tilde{f}(A (p\partial)\Phi_+ + (1-A)(p\partial)\Phi_-)
\end{equation}
Once again we encounter the notorious parameter $A$ resulting from
$\theta\cdot\delta\approx A\delta$. The (complex) Klein-Gordon equation
gives rise to a conserved current $j_{a}=(1/i)(\bar{\Phi}D_{a}\Phi- D_{a}\bar{\Phi}\Phi),\,\, D_{a}\Phi=(\partial_{a}-ieA_{a})\Phi,\,\, D_{a}\bar{\Phi}=(\partial_{a}+ieA_{a})\bar{\Phi}$
for smooth initial data. Since current conservation may no longer
be deduced from the singular equation of motion, we will require it
to hold separately.
\begin{eqnarray}
j^a & \approx & \frac{1}{i}\left((\bar{\Phi}_-\Phi_+-\bar{\Phi}_+\Phi_-)p^a\delta
 -2ie\tilde{f}(B\bar{\Phi}_+\Phi_++(A-B)(\bar{\Phi}_+\Phi_-+
 \bar{\Phi}_-\Phi_+)+\nonumber\right.\\
&&\left. (1-2A+B)\bar{\Phi}_-\Phi_-)p^a \delta\right) +\theta_+ j_+^a+\theta_-j_-^a\\
\partial\cdot j & \approx &
\frac{1}{i}((p\partial)(\bar{\Phi}_-\Phi_+-\bar{\Phi}_+\Phi_-)+
(\bar{\Phi}_+(p\partial)\Phi_+-(p\partial)\bar{\Phi}_+\Phi_+)\nonumber\\
&&-(\bar{\Phi}_-(p\partial)\Phi_--(p\partial)\bar{\Phi}_-\Phi_-)
-2ie\tilde{f}(p\partial)(B\bar{\Phi}_{+}\Phi_{+}+\nonumber\\
&&+(A-B)(\bar{\Phi}_{+}\Phi_{-}+\bar{\Phi}_{-}\Phi_{+})+(1-2A+B)\bar{\Phi}_{-}\Phi_{-}))
\delta\approx 0
\end{eqnarray}
This entails
\begin{multline}
  \label{eq:cckg}
((p\partial)(\bar{\Phi}_-\Phi_+-\bar{\Phi}_+\Phi_-)+
(\bar{\Phi}_+(p\partial)\Phi_+-(p\partial)\bar{\Phi}_+\Phi_+)-\\
-(\bar{\Phi}_-(p\partial)\Phi_--(p\partial)\bar{\Phi}_-\Phi_-)
-2ie\tilde{f}(p\partial)(B\bar{\Phi}_{+}\Phi_{+}+\\
+(A-B)(\bar{\Phi}_{+}\Phi_{-}+\bar{\Phi}_{-}\Phi_{+})+(1-2A+B)\bar{\Phi}_{-}\Phi_{-}))=0
\end{multline}
The expression for the current $j^a$ contains an additional arbitrary
constant $B$, which arises from the non-linear relation
$\theta^2\cdot\delta\approx B\delta$.
Its appearance seems to completely spoil our strategy to use the conservation
law to fix the value of $A$, since it seems we now need another equation to
determine the value of $B$. However, this conclusion is premature, since due
to local nature of the conditon (\ref{eq:cckg}) we actually have an infinite
number of conditions for $A$ and $B$ and thus an overdetermined system.
Re-arranging the first three in (\ref{eq:cckg}) in the form
containing expression involving the left-hand-side of (\ref{eq:jckg}) and its
complex-conjugate, which upon inserting the corresponding right-hand-side
yields after simply comparing coefficients $A=1/2$ and $B=1/4$.
This is actually a non-trivial statement
according to the above mentioned (infinite) over-determinacy, which
shows that the (naive) guess
that $\delta=\theta_+'$, which would
reproduce $A=1/2$ in all the previous cases is inconsistent with
the required value for $B$. So as a bonus from current conservation
the expression for the current $j^a$ is associated to
\begin{equation}
  \label{eq:fckg}
j^a\approx\theta_+j_+^a+\theta_-j_-^a +
  \delta\bar{\Phi}_-\Phi_-\frac{e\tilde{f}}{1+\frac{e^2}{4}\tilde{f}^2}p^a,
\end{equation}
which contains an extra piece streaming tangential to the hyperplane
of the pulse along its generators. 
Thus we finally obtain for the junction conditions
\begin{equation}
  \label{eq:jckgf}
  (p\partial)\Phi_+=\frac{1-i\frac{e}{2}\tilde{f}}{1+i\frac{e}{2}\tilde{f}} (p\partial)\Phi_-
\end{equation}
which once again take the form of a Cayley-transform. Note that the matching conditions are not sufficient to determine the solution uniquely. To solve the characteristic initial value problem would require additional boundary condition along a $(\bar{p}x)=v=const. $ surface.

\section{Dirac particle}

We will now consider a charged particle with spin  one-half subject to an
impulsive electromagnetic field. We therefore turn to Dirac's equation, which
is written in two-spinor form 
\begin{align}
  \label{eq:di}
  &\hat{P}_{AA'}\psi^A=\frac{m}{\sqrt{2}}\chi_{A'} &\text{where}\quad
       \hat{P}_{AA'}=\frac{1}{i}\nabla_{AA'}\\
  &\hat{P}^{AA'}\chi_{A'}=\frac{m}{\sqrt{2}}\psi^A &\text{and}\quad
       \nabla_{AA'}=\partial_{AA'}-ieA_{AA'}
\end{align}
For he specific form of the potential (\ref{empot}) this becomes
\begin{align}
  \label{eq:dipp}
  &\partial_{AA'}\psi^A-i\frac{m}{\sqrt{2}}\chi_{A'}=ief o_{A'}o_A\psi^A\nonumber\\
  &\partial^{AA'}\chi_{A'}-i\frac{m}{\sqrt{2}}\psi^A=ief
  o^Ao^{A'}\chi_{A'}\quad
  \text{where}\quad p^a=p^{AA'}=o^Ao^{A'}
\end{align}
For an impulsive profile $f=\tilde{f}\delta(px)$ and the decomposition of
$\psi^A=\theta_+\psi_+^A+\theta_-\psi_-^A$ and
$\chi^{A'}=\theta_+\chi^{A'}_++\theta_-\chi^{A'}_-$
into solutions of the free equation above and below the pulse respectively, we find
\begin{align}
  &\delta(px)(o_{A'}(o_A\psi^A_+-o_A\psi^A_-) 
      -ie \tilde{f}(A o_A\psi^A_+ + (1-A)o_A\psi^A_-))\approx 0\nonumber\\
  &\delta(px)(o^A(o^{A'}\chi_{A'+}-o^{A'}\chi_{A'-}) 
      -ie \tilde{f}(A o^{A'}\chi_{A'+} + (1-A)o^{A'}\chi_{A'-}))\approx 0
\end{align}
which in turn yields
\begin{align}
\label{eq:diju}
  &o_A\psi^A_+-o_A\psi^A_-=-ie \tilde{f}
     (A o_A\psi^A_+ +(1-A)o_A\psi^A_-)\nonumber\\
  &o_{A'}\chi^{A'}_+-o_{A'}\chi^{A'}_-=-ie \tilde{f}(A o_{A'}\chi^{A'}_+ + (1-A)o_{A'}\chi^{A'}_-)
\end{align}
Once again we encounter the ``ambiguity'' $A$ arising from the product of
$\delta$ with $\theta_+$. As has been our strategy in the previous paragraphs
we invoke the conservation law of the Dirac-current
\begin{equation}
  \label{eq:dicu}
  \partial_aJ^a=0\qquad J^a=\psi^A\bar{\psi}^{A'}+\chi^{A'}\bar{\chi}^A
\end{equation}
which in the smooth context is a direct consequence of (\ref{eq:di}). The
conservation requirement is equivalent to

\begin{multline}
  \delta(px)\left((o_A\psi^A_+)(o_{A'}\bar{\psi}^{A'}_+)-(o_A\psi^A_-)(o_{A'}\bar{\psi}^{A'}_-)+\right.\\
  \left.(o_A\chi^A_+)(o_{A'}\bar{\chi}^{A'}_+)-(o_A\chi^A_-)(o_{A'}\bar{\chi}^{A'}_-)\right)\approx0
\end{multline}
Re-arranging terms and using (\ref{eq:diju}) we find, not unexpectedly,
$A=1/2$, which in turn yields for the junction conditions at $px)=0$
\begin{align}
  \label{eq:jDi}
  &o_A\psi_+^A=\frac{1-i\frac{e}{2}\tilde{f}}{1+i\frac{e}{2}\tilde{f}}\, o_A\psi_-^A,
       \nonumber\\
  &o_A\chi_+^A=\frac{1-i\frac{e}{2}\tilde{f}}{1+i\frac{e}{2}\tilde{f}}\, o_A\chi_-^A.
\end{align}
Note that in contrast to the Klein-Gordon particle no further
constant appears. The current itself is simply associated to its classical
parts above and below the pulse. In this regard the Dirac-particle is simpler
than its Klein-Gordon analogue.

\section{Conclusion}
We have considered the scattering of charged classical and quantum  particles
by impulsive electromagnetic waves. The  problem is reduced to the matching of
free solutions above and below the pulse.  The theory of Colombeau generalized
functions was applied to give a meaning to products of singular terms, however
leading to undetermined constants. We have shown that by implementing
conservation laws that follow automatically from the equation of motion for
smooth solutions, allows one to determine these constants. As examples we
discussed the scattering of relativistic charged point particles and their
quantum analogues i.e. charged Klein-Gordon and Dirac fields. In all cases we
obtained a unique scattering amplitude.   
A natural step further is to extend our approach to particles with
vectorial charge structure as well as gravity.
We think that this method can be applied to similar physical situations, whenever the scattering source can be modeled by impulsive waves.

\begin{figure}[h]
 \includegraphics[scale=0.8]{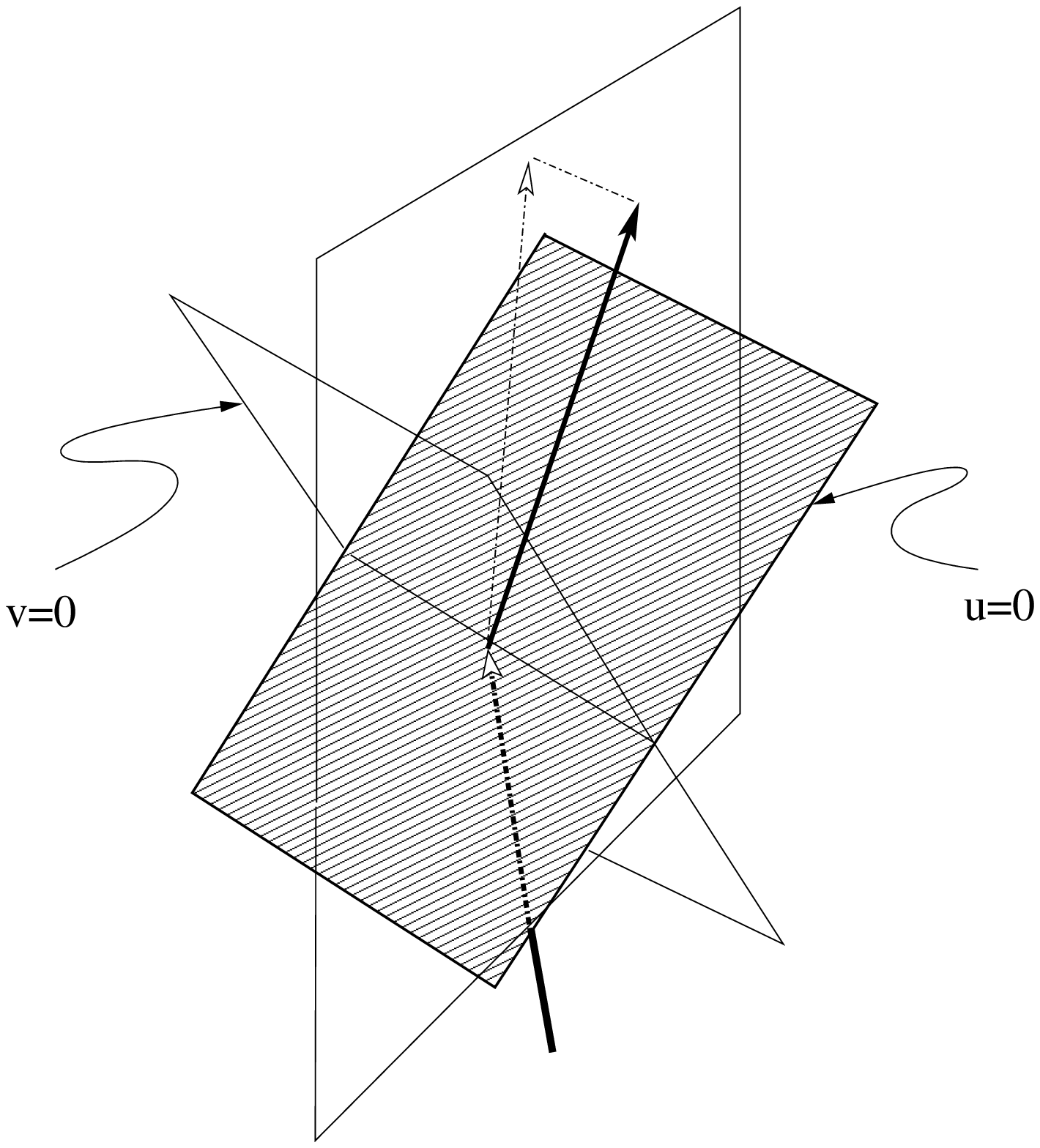} 
  \caption{The figure shows, schematically, the scattering of an incident
    particle with $\tilde{x}^i=0$ by an impulsive electromagnetic field
    supported at $u=0$. Notice that the particle is not only scattered within
    its plane of incidence but also off it}
  \label{fig:streu}
\end{figure}

\newpage
\begin{thebibliography}{99}
  \bibitem{Col}Colombeau J F, Multiplication of Distributions, LNM 1532,\\
    Springer 1992
  \bibitem{dVSa}DeVega H J and Sanchez N, Nucl.Phys.B {\bf 317}, 731 (1989)
  \bibitem{LoSa}Luosto C and Sanchez N, Nucl. Phys. B {\bf 355}, 231 (1991)
  \bibitem{AS}Aichelburg P and Sexl R, Gen. Relat. Gravit. 2: 303 (1971)
  \bibitem{RoLu}Rodnianski I and Luk J, arXiv:1209.1130, (2014) 
  \bibitem{KuSt}Kunzinger M and Steinbauer R, J.Math.Phys. 40 1479 (1999)
  \bibitem{Ba1}Balasin H, Class. Quantum Grav. 14 455 (1997)
  \bibitem{Pen}Penrose R, Ann.Phys. {\bf 10} 171, (1960) 
\end{thebibliography}
\end{document}